# Strategies to Maintain Voltage on Long, Lightly Loaded Feeders with Widespread Residential Level 2 Plug-in Electric Vehicle Charging

Don Scoffield
Mobility Systems & Analytics
Idaho National Laboratory
Idaho Falls, ID, USA
ORCID: 0000-0002-5650-0616
don.scoffield@inl.gov

John Smart
Mobility Systems & Analytics
Idaho National Laboratory
Idaho Falls, ID, USA
ORCID: 0000-0002-6648-9545
john.smart@inl.gov

Timothy Pennington
Infrastructure & Energy Storage
Idaho National Laboratory
Idaho Falls, ID, USA
ORCID: 0000-0002-4809-702X
timothy.pennington@inl.gov

C. Birk Jones
Renewable Energy &
Distributed Systems Integration
Sandia National Laboratories
Albuquerque, NM, USA
cbjones@sandia.gov

Matthew Lave
Renewable Energy &
Distributed Systems Integration
Sandia National Laboratories
Albuquerque, NM, USA
mlave@sandia.gov

Anudeep Medam
Energy Systems
Idaho National Laboratory
Idaho Falls, ID, USA
ORCID: 0000-0002-6648-9545
anudeep.medam@inl.gov

Bhaskar Mitra
Energy Systems
Idaho National Laboratory
Idaho Falls, ID, USA
ORCID: 0000-0001-5584-1390
bhaskar.mitra@inl.gov

*Abstract*—Long, lightly loaded feeders serving residential loads may begin to experience voltage excursions as plug-in electric vehicle (PEV) penetration increases. Residential PEV charging tends to occur during peak-load hours on residential feeders, leading to increased peak loads and potential voltage excursions. To avoid voltage excursions, two PEV charging control strategies were investigated using the IEEE 34-bus feeder. The first strategy shifts PEV charging energy from peak hours to off-peak hours; the other strategy allows PEVs to provide reactive power support. Undervoltage excursions seen in a simulation of uncontrolled charging of 200 PEVs were improved dramatically when these two control strategies were used. The minimum voltage on the feeder improved from 0.855 pu when PEV charging was uncontrolled to 0.959 pu when both control strategies were applied together.

*Keywords—maintaining voltage, PEV charging stations, electric utilities, optimization strategies*

## I. INTRODUCTION

With automakers announcing billions of dollars of investment in plug-in electric vehicle (PEV) development and multiple countries and U.S. states planning to prohibit the sale of internal-combustion-engine-vehicles in the future, PEV market penetration is expected to significantly increase [1-4]. To supply the future PEV market with power, one study suggests the U.S needs to deploy approximately 330,000 publicly available PEV charging stations by 2025 [5]. The present power grid infrastructure was not designed to accommodate this projected influx of plug-in electric vehicles (PEV) and charging stations [6]. Electric utilities need to develop strategies to mitigate the impact of such growth in PEV charging demand.

The simplest charging infrastructure are the residential single-phase, plug-in systems, which are categorized as alternating current (AC) Level 1 and Level 2 electric vehicle supply equipment (EVSE) that charge at 120 V and 240 V AC respectively [7]. The integration of a large number of PEVs charging at the same time may lead to large increases in distribution feeder peak loads and unacceptable voltage variations [8,9]. Integrating PEV loads without overloading the distribution system and managing PEV loads to minimize voltage variations pose an interesting challenge that must be solved to enable widespread PEV adoption. A promising approach to solve this challenge is to control PEV charging [10].

In this paper, the authors utilize an agent-based modeling and simulation platform to investigate: (1) voltage excursions that uncontrolled PEV charging can have on a long, lightly loaded feeder system; and (2) the ability of smart charge management to mitigate these voltage excursions by altering PEV charging loads.

## II. METHODOLOGY

### A. Problem Statement

Long, lightly loaded feeders tend to have voltage excursions as feeder loads increase, requiring mitigation actions provided by capacitor banks and tap-changing transformers.

Uncontrolled, home-based, AC Level 2 charging of PEVs tends to increase the residential peak loads because PEV charging occurs during the existing residential peak load times. This begs the question: if there is widespread Level 2 charging on a long, lightly loaded feeder system serving residential load, what will the extent of voltage excursions be? Can the voltage excursions be mitigated by smart PEV charge management or will more traditional grid mitigations need to be employed, such as upgrading feeder lines, capacitor banks and tap-changing transformers?

### B. Caldera™ Simulation Platform

The authors addressed these questions through modeling and simulation using Caldera™, a PEV charging infrastructure simulation platform designed to study the impact of PEV charging on the grid and develop strategies to manage charging. Its foundation is a library of high-fidelity PEV charging models derived from extensive charging and battery testing data that Idaho National Laboratory (INL) has collected over the past decade. Caldera's charging models accurately estimate charge power profiles, efficiency, and power factors for a wide variety of PEVs and charging technologies under varying grid conditions. These charging models are implemented through an agent-based modeling approach representing each PEV as an individual load on the distribution system. Caldera also contains smart charging algorithms for studying how to efficiently manage charging. Caldera also has modules to explore how artificial intelligence can help drivers and automated vehicles make intelligent charging decisions. Caldera enables the co-simulation of the transportation network and the grid. By linking existing simulation tools with Caldera, the grid impact of PEV charging can be accurately modeled.

### C. Electric Vehicle (EV) Charging Controls

The two smart charge management strategies investigated in this work employ an incentive-based control that would eliminate the need for direct communication with an aggregator. The first control strategy involves shifting the PEV charge energy from peak-load hours to off-peak-load hours, which is referred to in this paper as "Energy Shifting." The second control strategy assumes that PEVs can provide reactive power support, and is hereafter referred to in this paper as "Reactive Power Support." Each of these control strategies will be described in turn in the sections that follow.

#### 1) Energy Shifting Control Strategy

The Energy Shifting control strategy shifts much of the PEV charging energy to off-peak hours by randomizing the charge start time during the period when the vehicle is parked and plugged in, while ensuring there is sufficient time for the PEV to charge before it is needed again for driving. Specifically, the charge start time is selected from a random variable that is uniformly distributed on the interval from the park start time to the park end time minus the time required to fully charge.

#### 2) Reactive Power Support Control Strategy

The Reactive Power Support control strategy assumes that PEVs can provide reactive power. The real power (P) and reactive power (Q) generated by the PEV charger are constrained based on the apparent power (S) rating of the charger according to equation (1).

$$S^2 = P^2 + Q^2 \qquad (1)$$

The Reactive Power Support control strategy requires that PEVs charge at a reduced real power charge rate, in order to free up capacity to be used for reactive power support. In this paper, when the Reactive Power Support strategy is used, the PEVs real power charge rate is set to 70% of the apparent power rating of the charger, thereby allowing reactive power support between +71.4% and -71.4% of the apparent power rating.

The reactive power setpoint is determined from the node voltage using a volt-var curve. The volt-var curve maps the node voltage where the PEV is charging to a reactive power setpoint. The volt-var curve used in this paper is shown in Fig 1.

### III. CASE STUDY

#### A. Setup

A simulation case study was conducted to explore the charging impacts of 200 PEVs on the IEEE 34-bus feeder, which is a canonical example of a long, lightly loaded feeder. The locations of the PEVs on the feeder are shown in Fig. 2. The PEVs at each of the locations were connected to the same phase (e.g., phase 2).

The non-PEV load used in the simulation is a scaled version of a typical Pacific Gas & Electric (PG&E) residential load profile, downloaded from the PG&E website with a peak load of just over 3 MW [11].

All PEV charge models assumed a 50-kWh battery pack and an 6.6-kVA Level 2 on-board charger. The charging behavior of the PEVs was derived from real-world charging of PEVs in the PG&E service territory collected in the EV Project by INL [12]. The charging behavior is characterized by PEV arrival time, departure time, arrival state of charge (SOC) of the vehicles' batteries, and battery SOC at the end of the charge, as requested by the drivers.

The assessment presented here describes the impact of uncontrolled and controlled PEV charging on the feeder's peak, overall voltages, and the worst case voltages. The paper quantifies the change in the feeder's load profile when subjected to the different controls. Then, the paper reviews the voltage response to the change in the feeder's loading under the different scenarios.

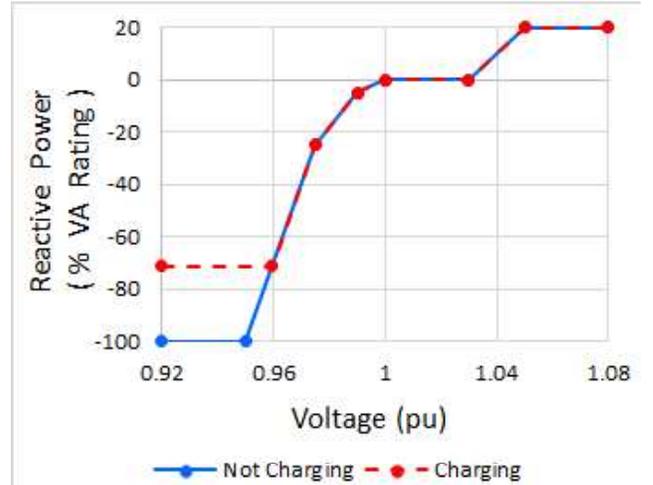

Fig. 1. Volt-var curve used in the Reactive Power Support control strategy. The sign convention views PEV as a load and not a generator, negative reactvie power is capacitive load that supplies Q and positive reactive power is inductive load that consumes Q.

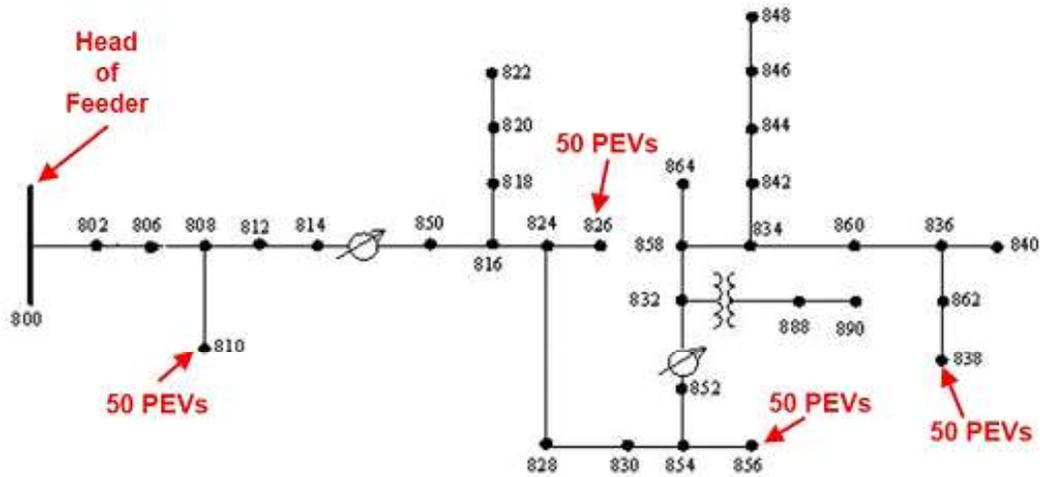

Fig. 2. IEEE 34 bus feeder, which is a canonical example of a long lightly loaded feeder. Fifty PEVs charge on nodes 810.2, 826.2, 856.2, and 838.2. Node 856.2 is used for comparing scenarios.

*B. Uncontrolled PEV Charging*

On the IEEE 34-bus feeder, the uncontrolled charging of 200 PEVs caused the feeder peak load to increase by a significant amount from the baseline, where no PEVs were charging, as shown in Fig. 3(a). The uncontrolled peak feeder demand at 19:29, indicated by the black dot in Fig. 3(a), was at 3.68 MW, which was 19% higher than the no-PEV scenario's max load. A spatial representation of the peak loading at this time on the distribution electric power system, as depicted in Fig. 3(b), describes the magnitude of the PEV loads in comparison to the non-PEV residential loads. The gray marker sizes in Fig. 3(b) are proportional to the power demand at each node on the grid. At the four PEV charging locations, indicated by the yellow squares, the vehicles' demand clearly exceeded the non-PEV residential load. Also, each of the four PEV loads exceeded all of the non-PEV residential loads scattered throughout the system, except for the load at node 844.

Higher-power demands attributed to PEV charging caused substantial changes in node voltages to occur. When there was no PEV charging, voltages for all nodes were between 0.95 and 1.05 p.u., as seen in Fig. 4. As expected, an increase in demand resulted in lower voltages; the significant decrease in voltage from the no-PEV case to the uncontrolled-PEV simulation is evident in Fig. 4. In the uncontrolled-PEV scenario, the lowest node voltage dropped below the American National Standards Institute (ANSI)-defined threshold of 0.95 p.u to only 0.85 p.u.

The undesirable minimum voltage typically caused by an increase in load could also be attributed to an imbalance on the distribution system. The simulations added PEV charging loads to different locations on the same phase. This resulted in a load increase of greater than 150% on one phase and no change to the other two. This represents an unlikely but worst-case scenario that the utility would have to overcome to maintain appropriate voltages throughout the system. The next section looks at using PEV controls to maintain voltages throughout the system for this worst-case scenario.

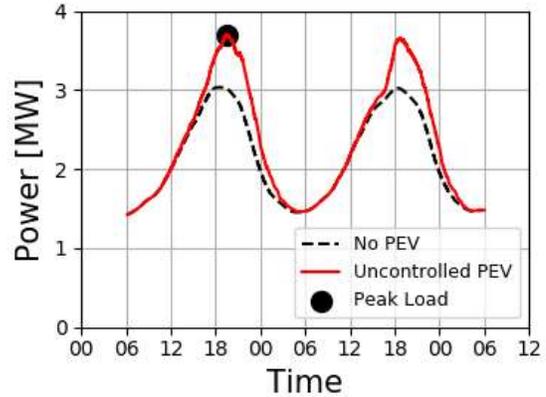

(a) Timeseries feeder power over two days without PEV charging and with uncontrolled PEV charging. The feeder peak with uncontrolled PEV load occurred at 19:29 on the first day.

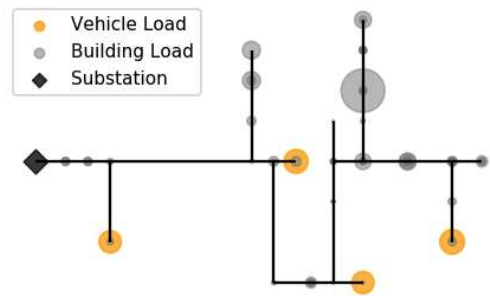

(b) PEV and non-PEV load magnitudes at uncontrolled PEV feeder peak. The magnitudes of the markers are proportional to the load.

Fig. 3. Uncontrolled PEV charging resulted in a noticeable increase in feeder load. (a) Two scenarios (no PEV and uncontrolled PEV) plotted over a two-day period describe the increase in load caused by PEV charging. At the peak load, PEV charging resulted in an increase of 19%. (b) At feeder peak, the topology map shows the magnitude of the PEV and non-PEV loads.

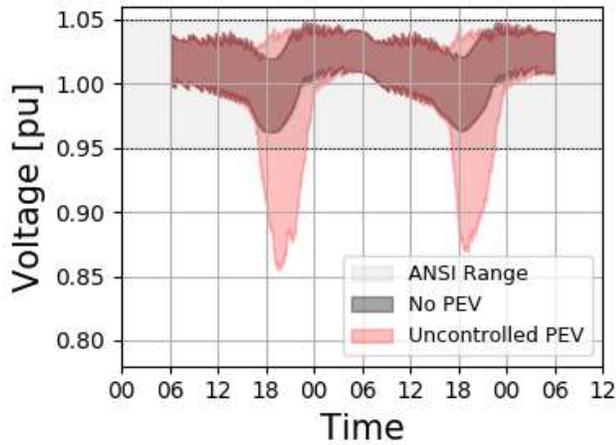

Fig. 4. When subjected to only non-PEV loads, the voltages of the feeders remained between the ANSI voltage threshold. However, PEV uncontrolled charging added significant load to the phase 2 lines, which caused the minimum voltage to drop well below 0.95 p.u. and reach a minimum of 0.85 p.u.

*C. Controlled PEV Charging*

To mitigate consequences associated with an increased load due to vehicle charging, this work found that shifting PEV loads and providing reactive power support limits the drop in voltage seen in the uncontrolled case. The simulations found that including a combination of energy shifting and reactive power support results in the least amount of voltage drop.

The energy-shifting approach attempted to improve operations by limiting active power at the peak. This is evident in Fig. 5(a) where the feeder power profile under energy-shifting controls results in only a slight increase in the peak; the feeder peak reached a high of 3.15 MW, an 3.8% increase over the no-PEV peak of 3.03 MW, whereas the uncontrolled PEV case had a peak of 3.68 MW. The reduction in the peak load compared to the uncontrolled case resulted in more energy consumption during the night time hours between 18:00 and 11:00 the next day. In contrast, the reactive power control approach did not change the feeder power profile when compared with the uncontrolled PEV scenario.

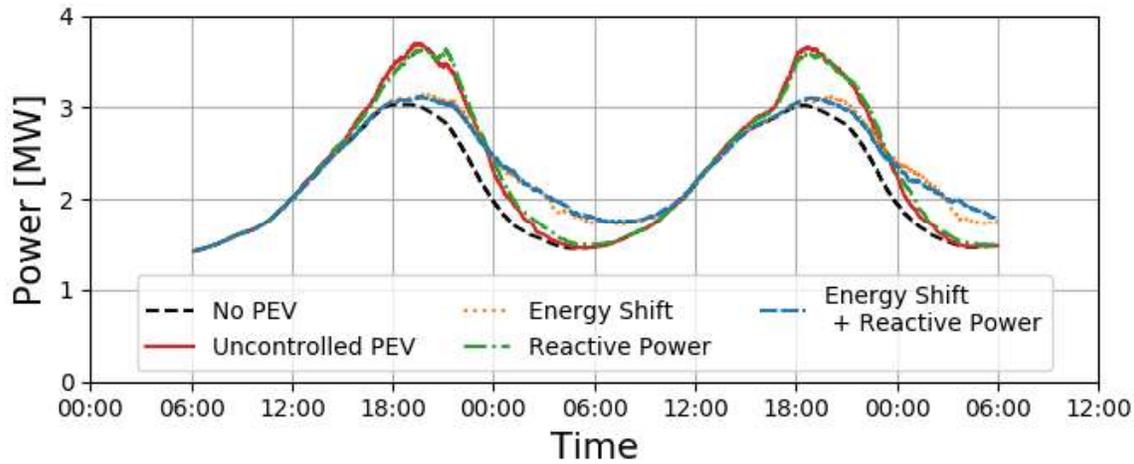

(a) The feeder profiles over a two-day period in the five simulation scenarios had similarities and differences depending on the control approach. The reactive-power control approach matched well with the uncontrolled case because very little change in active power occurred. The energy-shift and energy-shift + reactive-power approaches had a similar peak with the no-PEV case, but had larger power draws during the night.

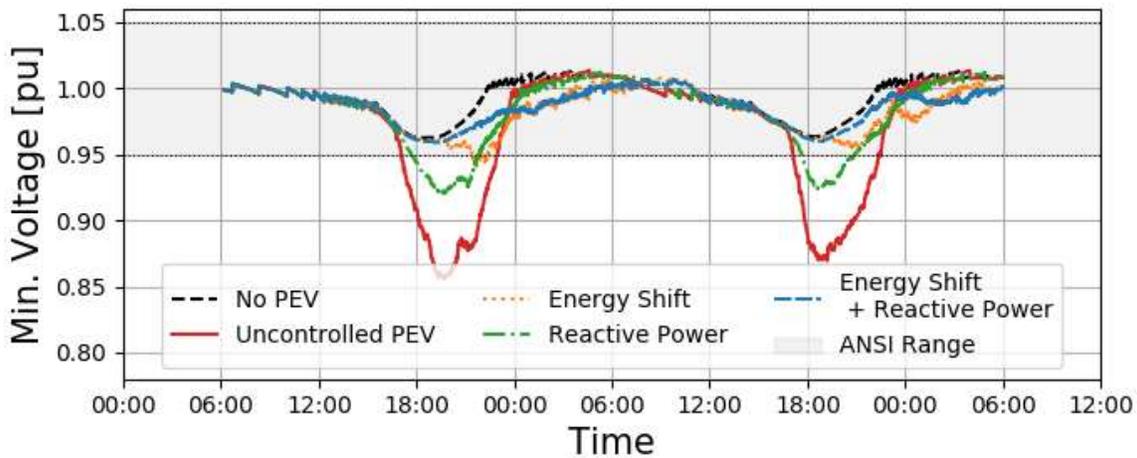

(b) The impelementation of PEV controls had different impacts on the feeder minimum voltages over the two-day period. The reactive-power-control approach alone provided the least amount of benefits and did not eliminate volations where the minimum voltage dropped below 0.95 p.u. The energy-shift method performed better, but still saw a brief instance where the voltage dropped below the threshold. However, when combined, the energy-shift and reactive-power-control approaches together provided sufficient voltages that often matched the no-PEV case.

Fig. 5. (a) Feeder power profiles and (b) minimum voltages show how the electric power system changed and responded to PEV charging controls. .

Allowing the PEV charging stations to provide reactive power improved the voltage by injecting reactive power onto the electric power system. This control action did not cause significant changes to occur in the active power, as depicted by the green line in Fig. 5(a). However, when the reactive power capability was added to the energy-shifting control mechanism, the feeder profile experienced a change in the profile that decreased the peak, as shown by the blue line in Fig 5(a).

The control methods used here provided some relief to the system by increasing the voltage above the uncontrolled case, as shown in Fig. 5(b). When PEVs were charging, the voltages remained above the 0.95 p.u. threshold only when the energy-shift and reactive-power control approaches were used together. On its own, the energy-shifting method experienced a drop in voltage on the first day that fell below the ANSI threshold. When using the energy-shifting control approach, the voltage matched the no-PEV case until 19:41, at which time it began to decrease and eventually dropped below the ANSI threshold of 0.95 p.u when the feeder load peaked.

Implementing only the Reactive Power Support control by enabling the charging stations to inject reactive power at low voltage provided a noticeable change from the uncontrolled case, but was not enough to avoid ANSI voltage violations. To provide sufficient support, the reactive power control method had to be combined with the energy shifting approach. In this case, where both reactive power and energy-shifting were enabled, the voltage measurements had similar results with the no-PEV scenario and did not experience ANSI violations.

### D. PEV Charging Control Overview

The IEEE 34-bus feeder experienced a significant drop in voltage when subjected to PEV loads on a single-phase that were mitigated at different levels by the control approaches. Fig. 6 depicts the electric power system voltages during on- and off-peak operations. This paper deemed on-peak operations to be between 16:00 and midnight when the voltage deviated farthest from the nominal voltage of 1.0 p.u. Off-peak operations included all other hours measured over the two-day period. The boxplots clearly show that when both the energy-shifting and reactive power control approaches were used together, the system voltages matched closely with the no-PEV case. Energy shifting and reactive power alone had larger deviations from the no-PEV scenario and had instances where voltage dropped below 0.95 p.u. Finally, during off-peak operations, all of the methods corresponded well with the no-PEV simulation case.

## IV. CONCLUSIONS AND FUTURE WORK

### A. Conclusions

Long, lightly loaded feeders serving residential loads may begin to experience voltage excursions as the number of PEVs on the feeder increases. These voltage excursions are caused by PEVs charging during peak times, which increases feeder peak load. Voltage excursions would normally be mitigated by upgrades to the feeder system or with tap-changing transformers. This paper investigated the potential for PEV charging control strategies to mitigate voltage excursions without grid upgrades. Two PEV charging control strategies were investigated with Caldera using the IEEE-34 bus feeder. The undervoltage excursions caused by uncontrolled PEV charging were significantly improved by the two control strategies. The minimum voltage on the feeder improved from 0.855 pu when PEV charging was uncontrolled to 0.959 pu when both control strategies were applied together. The first strategy shifted PEV charging energy from peak hours to off-peak hours; the other strategy allowed PEVs to provided reactive power support. Because these strategies use different mechanisms to support voltage, they can be employed independently and simultaneously. When used together, these strategies mitigated the voltage excursions caused by uncontrolled PEV charging so that all node voltages were above the ANSI-defined threshold of 0.95 p.u.

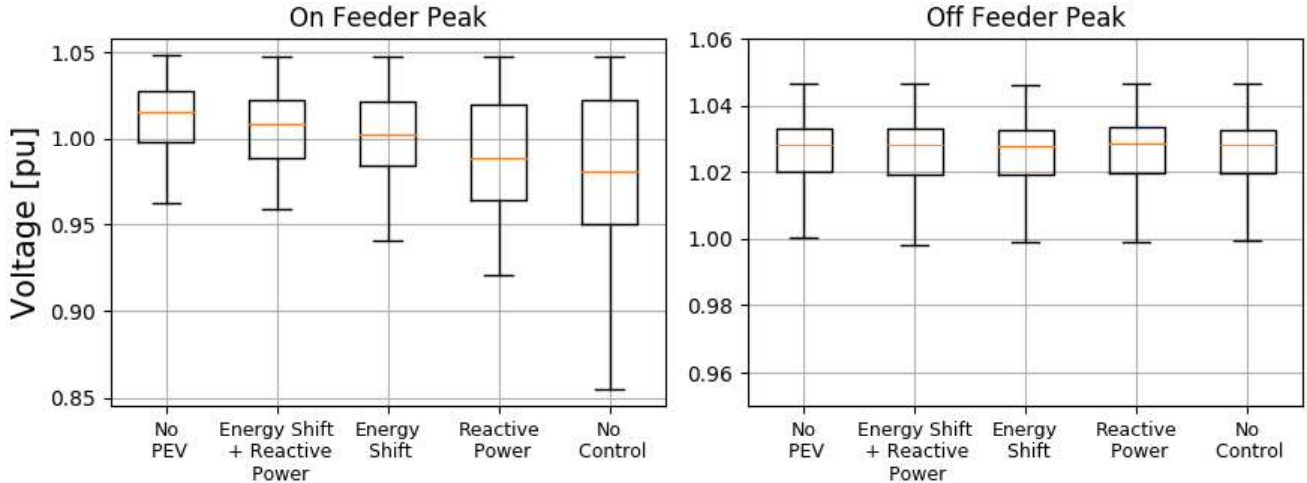

Fig. 6. Boxplot of feeder voltages for the five simulation scenarios during on-peak (hour 16-midnight) and off-peak (hour 0-15) time periods explain the overall differnce between the approaches. Each approach improved upon the no control situation, but the energy-shift and reactive power methods by themselves still resulted in voltage violations. But, using the energy-shift approach in conjunction with the reactive power support approach resulted in voltages that matched closely with the no PEV scenario.

## B. Future Work

An important area of future work is to compare the cost/benefit of PEV charging controls described here with traditional feeder upgrades and more advanced control strategies. This comparison should compare the grid benefits offered by each option to the cost to implement and maintain the approaches. For example, it is very likely that state-of-the art controls will outperform the controls studied here, but the more advanced controls will almost certainly cost more to implement and maintain. Understanding whether the additional benefit is worth the additional cost is a necessity for efficient implementation of PEV charging control.

ACKNOWLEDGMENTS

This work is made possible due to funding of the U.S. Department of Energy Vehicle Technologies Office. This information was prepared as an account of work sponsored by an agency of the U.S. Government. Neither the U.S. Government nor any agency thereof, nor any of their employees, makes any warranty, express or implied, or assumes any legal liability or responsibility for the accuracy, completeness, or usefulness of any information, apparatus, product, or process disclosed, or represents that its use would not infringe privately owned rights. References herein to any specific commercial product, process, or service by trade name, trademark, manufacturer, or otherwise, does not necessarily constitute or imply its endorsement, recommendation, or favoring by the U.S. Government or any agency thereof. The views and opinions of authors expressed herein do not necessarily state or reflect those of the U.S. Government or any agency thereof.